\documentclass[11pt]{article}
\usepackage{graphicx}
\usepackage{amsmath}
\usepackage{amsfonts}
\usepackage{amssymb}
\usepackage{epsfig}

\newcommand{\be}{\begin{equation}}
\newcommand{\ee}{\end{equation}}

\newcommand{\ba}{\begin{eqnarray}}
\newcommand{\ea}{\end{eqnarray}}

\begin{document}

\title{How a cold axion background influences photons}
\author{{ Dom\`enec Espriu, Albert Renau}\\
\small Departament d'Estructura i Constituents de la Mat\`eria \\\small Institut de Ci\`encies del Cosmos (ICCUB),\\
\small Universitat de Barcelona, Mart\'\i ~i Franqu\`es 1, 08028 Barcelona, Catalonia, Spain\\
}

\date{}

\maketitle

\begin{abstract}
A cold relic axion condensate resulting from vacuum misalignment
in the early universe oscillates with a frequency $\sim m_a$, where $m_a$ is the
axion mass.
We summarize how the properties of photons propagating in such a medium are modified. Although
the effects are small due to the magnitude of the axion-photon coupling, some consequences are striking.
\end{abstract}

\vspace{-10cm}
\begin{flushright} ICCUB-11-203\\
UB-ECM-PF-64/11
\end{flushright}  
\vspace{8.5cm}

\section{Introduction}
Cold relic axions resulting from vacuum misalignment in the early universe is a valid candidate for
dark matter \cite{darkmatter}. In this model, a coherent spatially constant axion field (it may be a Peccei-Quinn
axion \cite{PQ} or a similar field) acquires a mass $m_a$ once
instanton effects set in to find itself, in general, not in the minimum of the potential.
In late times, the axion field oscillates,
\begin{equation}\label{cos} a(t)= a_0 \cos m_a t,
\end{equation}
and the energy density stored in these oscillations is $\rho\simeq a_0^2 m_a^2$, contributing to the energy budget.

Axions affect photons in a universal way, described by ${\cal L}_{a\gamma\gamma}$
\begin{equation}
\mathcal L_{a\gamma\gamma}=g_{a\gamma\gamma}\frac{\alpha}{2\pi}\frac a{f_a}F_{\mu\nu}\tilde F^{\mu\nu}.
\nonumber
\end{equation}
Popular models such as DFSZ  and KSVZ \cite{models} all give $g_{a\gamma\gamma}\simeq 1$. Using \eqref{cos} this becomes
\begin{equation}
\mathcal L_{a\gamma\gamma}= -  g_{a\gamma\gamma} \frac{\alpha}{\pi} \frac{a_0}{f_a}
\cos (m_a t)\, \epsilon^{ijk} A_i F_{jk}.
\nonumber
\end{equation}
 Note that the above term is Lorentz non-invariant.

A natural question is whether and how the presence of this axion background could be detected thus making the
cold axion hypothesis plausible. Three possible effects are proposed: (1) Cold axions influence cosmic ray
propagation and induce photon Bremsstrahlung; (2) Cold axions induce an additional rotation in the polarization plane of light
(on top of the familiar one \cite{raffelt}); and (3) Some photon wave-lengths are forbidden in a
universe filled with cold axions.

In this talk we shall concentrate mostly on points (2) and (3). Point (1) has been discussed in detail in \cite{ours}.
In order to determine the properties of photons in such a medium we need to solve the equation of motion in momentum space
\begin{equation} \left[g^{\,\lambda\nu}\left(k^2-m^2_\gamma\right) +
i \,\varepsilon^{\,\lambda\nu\alpha\beta}\,\eta_\alpha\,k_\beta\right]
\tilde A_\lambda(k)=0.
\label{movement}
\end{equation}
where $\eta_\alpha\sim  \partial_\alpha a= \delta_\alpha^0 \dot a$.
Two complex and space-like chiral polarization vectors
$\varepsilon^{\mu}_{\pm}(k)$ can be defined (see \cite{sasha}).

Let us now assume that we are dealing with photons of momenta
$|{\bf k}|\gg  m_a$. Then it makes sense to treat in (\ref{movement}) the axion background
adiabatically with
$\eta_\alpha=(\eta_0,0,0,0)$ where
$\eta_0$ can be taken as approximately constant, and $\eta_0=2g_{a\gamma\gamma}\frac\alpha\pi\frac{a_0m_a}{f_a}$.
 The two polarization vectors
are solutions of the vector field equations if and only if
\begin{equation} k^{\mu}_{\pm}=(\omega_{{\bf k}\,\pm} , {\bf k})\qquad
\omega_{{\bf k}\,\pm}=\displaystyle
\sqrt{{\bf k}^2+m_{\gamma}^{2}\pm\eta_0 |{\bf k}|}.
\label{reldis}\end{equation}
The astrophysical and observational bounds indicate that $ \eta_0 < 10^{-20} {\rm eV} $.
The situation when $\vert{\bf k}\vert\le m_a$ will be discussed later.

\section{Bremsstrahlung}

The process $p({\bf p}) \to p({\bf p}-{\bf k})\gamma({\bf k})$ (or $e \to e \gamma$) is possible in a
Lorentz-violating theory.
The cold axion background provides the necessary source of Lorentz non-invariance.
Cosmic rays can lose energy due to the radiation of photons while preserving momentum and energy conservation
thanks to eq. (\ref{reldis}).
The differential decay width for the process is \cite{ours}
\begin{equation} d\Gamma(Q)
=\frac\alpha2\frac{|{\bf k}|}{|{\bf p}|}\frac1{E\omega_{\bf k}}(-p\cdot k+|{\bf p}|^2\sin^2\theta)d|{\bf k}|,
\nonumber
\end{equation}
and in the relevant limit $E \ll m_p^2/|\eta_0|$ we have
\begin{equation} \frac{dE}{dx}=-\frac{\alpha\eta_0^2 E^2}{4m^2_p},\qquad E(x)=
\frac{E(0)}{1+\frac{\alpha\eta_0^2}{4m_p^2} E(0) x}.\nonumber
\end{equation}
From the likely detection of extragalactic cosmic rays we can get a model independent bound
$\eta_0 < 10^{-15}{\rm ~eV}\Rightarrow f_a > 100 {\rm ~GeV}$ (i.e. exclude weak scale axions in a
completely model independent way).

However the total amount of energy loss fos a cosmic ray is very low. It is therefore more meaningful to
look for the radiated photons. Those emitted from electrons have a spectrum which is different from the overwhelming
galactic synchrotron radiation background. Several hypothetical ways of detecting this radiation are discussed in
\cite{ours}.

\section{Exotic rotation of the polarization plane}

In order to find the effect of axions on the photon propagator one has to consider:

--- A constant magnetic field (well known result). For simplicity we shall assume ${\bf k}\cdot \vec B=0$.

--- The cold axion background (new).

\begin{figure}[h]
\center
\includegraphics[scale=0.25]{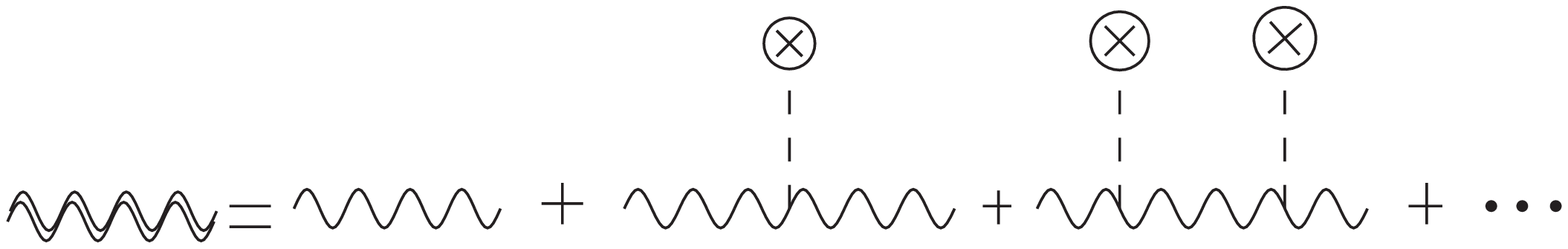}\hspace{1cm}
\includegraphics[scale=0.28]{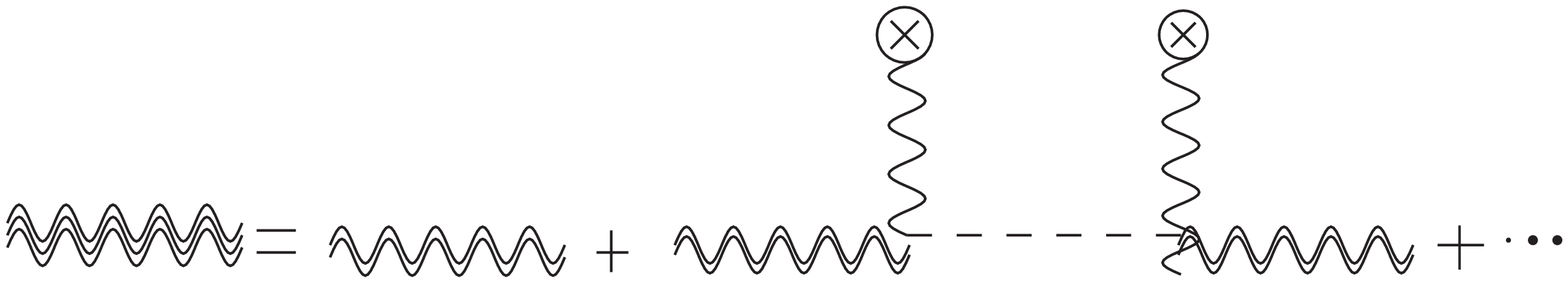}
\caption{Propagator in the cold axion background and the magnetic field}\end{figure}
The photon propagator for ${\bf k} \cdot \vec B=0$ is ($\vec b \equiv \frac{2g_{a\gamma\gamma}\alpha}{\pi f_a} \vec B$)
\begin{equation} \mathcal D_{\mu\nu}(k)\simeq \frac{-i g_{\mu\nu}}{k^2}+\frac{ik_0^2b_\mu b_\nu}{k^2[k^2(k^2-m^2)-k_0^2\vec b^2]}
-g_\mu^j g_\nu^l\frac{\eta_0 k_0^2\left[b_j(\vec b\times \vec k)_l-b_l(\vec b\times \vec k)_j\right]}
{k^4\left[k^2(k^2-m^2)-k_0^2\vec b^2 \right]}.\nonumber
\end{equation}
Then, for a photon plane wave initially with an electric field forming an angle $\beta$ with $\vec B$, the plane of
polarization rotates as
\begin{equation} \tan2\alpha(x)=\frac{[1+2f(x)]\sin2\beta+3\eta_0|x|\cos2\beta}{4f(x)+[1+4f(x)]\cos2\beta-3\eta_0 |x|\sin2\beta},\quad f(x)=\frac{\vec b^4}{16m^4}k_0^2|x|^2,\nonumber
\end{equation} and the expected value for the angle is
\begin{equation} \bar\alpha=-\frac12\frac{\left[1+2f(x)\right]\sin2\beta+3\eta_0|x|\cos2\beta}{[1+4f(x)]+4f(x)\cos2\beta}.\nonumber
\end{equation}
The rotation survives even without magnetic field and the effect is independent of the frequency.
Note that the previous results hold only for table-top experiments when the photon can approximately be considered
an eigenstate of energy, i.e. for a period where the time-of-flight of the photon is smaller than $2\pi /m_a$.
We have not considered the relevance of the effect when the latter condition is not fulfilled. It is clear, however, that
then the effect of the cold axion background has to be proportional to $\eta_0^2$ instead of $\eta_0$.

\section{What if $|{\bf k} | \le m_a$? Forbidden wavelengths}

\begin{figure}[h]
\center
\includegraphics[scale=0.7]{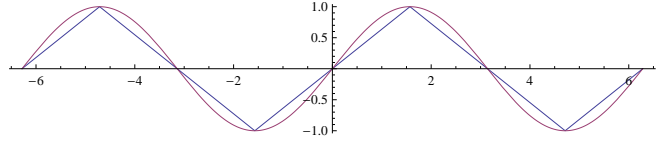}
\caption{Triangle wave versus sine}
\label{tsin}\end{figure}

As mentioned in the introduction,
$a(t)$ changes sign with a period $2\pi/m_a$ and this is now
relevant. Let us approximate the sinusoidal variation by a triangle wave (see \ref{tsin}) and solve
exactly for the propagating modes.
The equation for $\hat A_\nu(t,\vec k)$ is
\be\label{eq}
\left[g^{\mu\nu}(\partial_t^2+\vec k^2)-i\epsilon^{\mu\nu\alpha\beta}\eta_\alpha k_\beta\right]\hat A_\nu(t,\vec k)=0.
\qquad
\hat A_\nu(t,\vec k)=\sum_{\lambda=+,-}f_\lambda(t)\varepsilon_\nu(\vec k,\lambda)\nonumber.
\end{equation}
We now write $f(t)=e^{-i\omega t}g(t)$ and demand that $g(t)$ have the same periodicity as $\eta(t)=\eta_0\sin(m_at)$. This
requires
\begin{equation}
\cos(2\omega T)=\cos(\alpha T)\cos(\beta T)-\frac{\alpha^2+\beta^2}{2\alpha\beta}\sin(\alpha T)\sin(\beta T),
\quad T= \frac{\pi}{M_a},\quad \alpha,\beta=\sqrt{k^2\pm \eta_0 k}.
\nonumber
\end{equation}
For $\eta_0\equiv \eta_{\max} \ll m_a$ there is no relevant variation with respect to the
decay rate computed assuming $|{\bf k}| >> m_a$ is found, and
this can be understood intuitively.
However if $\eta_0/m_a$ grows there is a surprise; some values of $|{\bf k} |$ admit no solution for $\omega$
\begin{figure}[h]
\center
\includegraphics[scale=0.3]{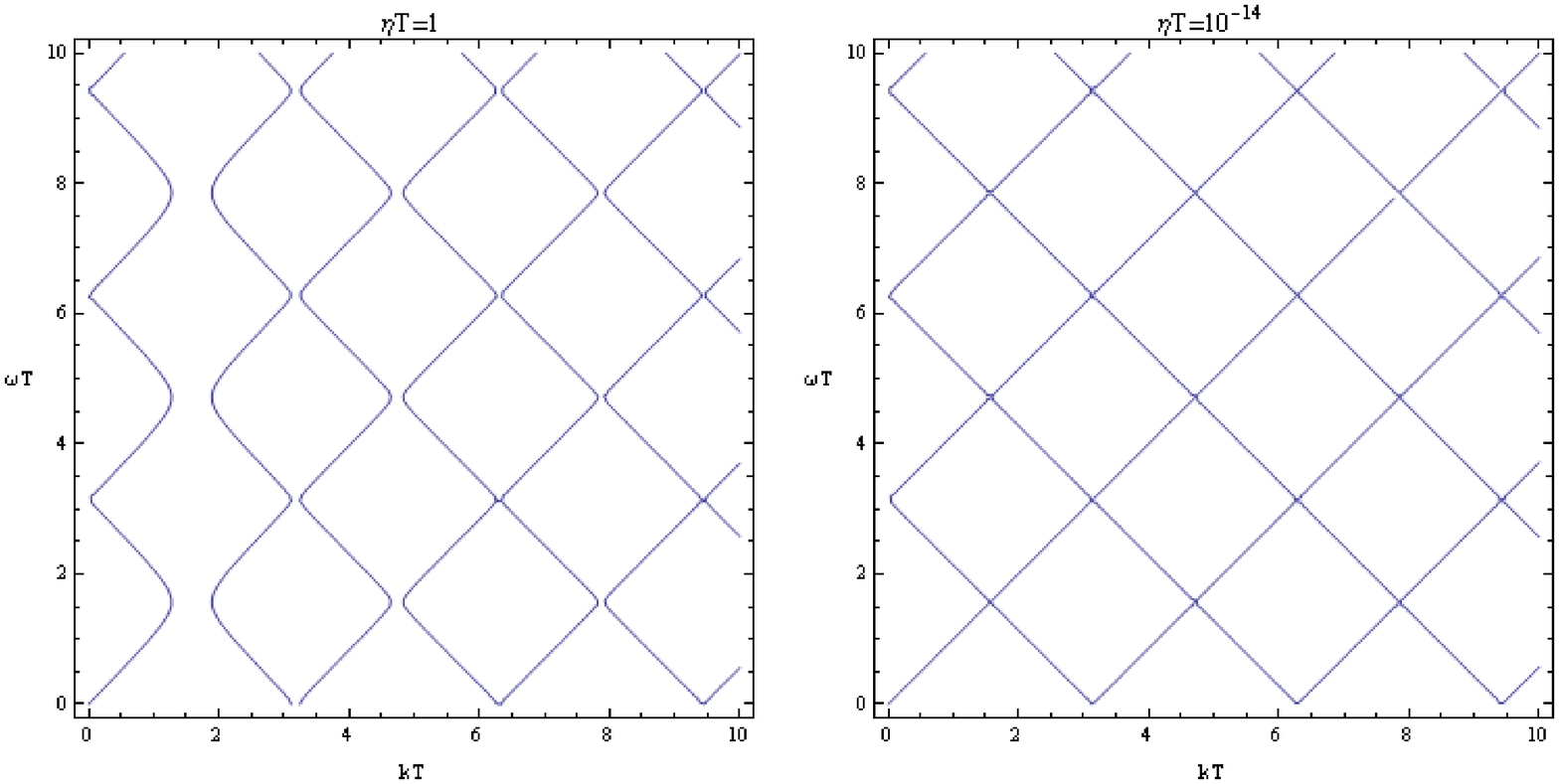}
\caption{Forbidden wavenumbers}
\end{figure}
We arrive then at the striking conclusion that in a universe filled with cold axions oscillating with a period
$2\pi/m_a$ some wavelengths are forbidden by a mechanism that is similar to the one preventing some energies from existing
in a semiconductor. Of course the width of the forbidden bands is very narrow (proportional to $\eta_0$),
hence difficult to detect.  Further details can be found in \cite{last}

\section*{Acknowledgments}

It is a pleasure to thank K. Zioutas and the rest of organizers of Patras 2011 for an ejoyable conference and encouragement.
We acknowledge financial support from projects FPA2010-20807, 2009SGR502 and Consolider CPAN.

\end{document}